\documentclass{PoS}

\usepackage{amsmath}
\DeclareMathOperator{\im}{Im}
\usepackage[]{graphicx}

\newcommand{\MS}{\ensuremath{\overline{\text{MS}}}}

\title{NNNLO determination of the bottom-quark mass from
non-relativistic sum rules}

\ShortTitle{NNNLO determination of the bottom-quark mass from
non-relativistic sum rules}

\author{Martin Beneke\\
        Physik Department T31, James-Franck-Stra{\ss}e 1, Technische
        Universit\"at M\"unchen,\\ 85748 Garching, Germany
        }

\author{\speaker{Andreas Maier}\,\footnote{Preprint numbers:
    SI-HEP-2016-01, TUM-HEP-1035/16}
\\
        IPPP, Department of Physics, University of Durham, DH1 3LE,
        United Kingdom\\
        \email{andreas.maier@durham.ac.uk}
        }

\author{Jan Piclum\\
        Albert Einstein Center for Fundamental Physics, Institute for
        Theoretical Physics,\\
        University of Bern, 3012 Bern, Switzerland\\
        Theoretische Physik 1, Naturwissenschaftlich-Technische
        Fakult\"at, Universit\"at Siegen, \\
        57068 Siegen, Germany
        }

\author{Thomas Rauh\\
        Physik Department T31, James-Franck-Stra{\ss}e 1, Technische Universit\"at M\"unchen,\\ 85748 Garching, Germany}

\abstract{
  The mass of the bottom quark can be determined with high precision
  from moments of the pair-production cross section $\sigma(e^+ e^- \to b
  \bar{b})$ near threshold. We present the first complete NNNLO
  determination from non-relativistic sum rules, obtaining a bottom-quark
  mass of \mbox{$m_b^\text{PS}(2\,\text{GeV}) = 4.532^{+0.013}_{-0.039}\,\text{GeV}$} in the
  potential-subtracted scheme. For the mass in the \MS{} scheme we find
  $m_b^{\MS} (m_b^{\MS}) = 4.203^{+0.016}_{-0.034}\,\text{GeV}$ using the recently
  computed four-loop correction to the scheme conversion.
}

\FullConference{12th International Symposium on Radiative Corrections (Radcor 2015) and LoopFest XIV (Radiative Corrections for the LHC and Future Colliders)\\
		 15-19 June  2015\\
		 UCLA Department of Physics \& Astronomy Los Angeles, CA, USA}

\begin{document}

\section{Introduction}
\label{sec:intro}

A precise knowledge of the bottom-quark mass is of considerable interest
not only by itself, but also for a number of phenomenological
applications like $B$ meson and Higgs decays.

For a precision determination it is crucial to find a quantity that
is both highly sensitive to the mass and well accessible by
experiment. This suggests considering the normalised production cross
section
\begin{equation}
  \label{eq:Rb_def}
  R_b(s) = \frac{\sigma(e^+e^- \to b\overline{b} + X)}{\sigma(e^+e^- \to
    \mu^+ \mu^-)}
\end{equation}
near threshold. In this region, $R_b$ is dominated by non-perturbative effects.
Within the sum rule approach~\cite{Novikov:1976tn,Novikov:1977dq}, however,
it has been argued that these effects largely cancel out when considering moments
\begin{equation}
\label{eq:Mn_def}
{\cal M}_n = \int_0^\infty ds\,\frac{R_b(s)}{s^{n+1}}\,.
\end{equation}
The bottom-quark mass can thus be determined by comparing the weighted
integrals over the experimentally measured cross section to the theory
prediction for the moments.

For sufficiently large values $n \gtrsim 10$ the moment integral in
eq.~\eqref{eq:Mn_def} is dominated by the threshold region $\sqrt{s}
\sim 2m_b$. In this region, the produced $b\bar{b}$ pair is
non-relativistic and the strong interaction leads to bound-state
formation and thus a breakdown of conventional perturbation theory. Both
of these phenomena are accounted for in the effective theory of potential
non-relativistic QCD (PNRQCD)~\cite{Pineda:1997bj}, where a simultaneous
expansion in $\alpha_s$ and the small quark velocity $v \sim 1/\sqrt{n}$
is performed. Contributions from Coulomb interaction scale as
$\alpha_s/v$ and are resummed to all orders. More specifically, the
power counting up to NNNLO is given by
\begin{equation}
  \label{eq:PNRQCD_power_counting}
  R_b \sim v \sum_k \bigg(\frac{\alpha_s}{v}\bigg)^{\!k}\times
  \begin{cases}
    1 & \text{LO}\\
    \alpha_s,v & \text{NLO}\\
    \alpha_s^2,\alpha_s v, v^2 & \text{NNLO}\\
    \alpha_s^3,\alpha_s^2 v,\alpha_s v^2, v^3 & \text{NNNLO}
  \end{cases}\,.
\end{equation}
To obtain reliable predictions it is necessary that the smallest scale
in the theory, given by the kinetic energy $E \sim m_b\*v^2 \sim m_b/n$,
remains above the typical scale $\Lambda_{\text{QCD}}$ of
non-perturbative physics.

\section{Determination of moments}

In the following we describe the determination of the experimental and
theory moments. In both cases we can split the moment integral into a
contribution from the narrow bound-state resonances and an integral over
the continuum cross section. For the moments ${\cal M}_n$ with $n\approx
10$
considered in this work the former contribution is dominant.

\subsection{Experimental moments}
\label{sec:M_exp}

Treating the four bound states $\Upsilon(1S)$ to $\Upsilon(4S)$ in the
narrow-width approximaton we obtain
\begin{equation}
  \label{eq:Mn_exp_res}
  {\cal M}_n^\text{exp} = 9\pi \sum_{N=1}^4
  \frac{1}{\alpha(M_{\Upsilon(NS)})^2}\frac{\Gamma_{\Upsilon(NS)\to l^+
      l^- }}{M_{\Upsilon(NS)}^{2n+1}}
  + \int_{s_\text{cont}}^\infty ds\, \frac{R_b(s)}{s^{n+1}}\,.
\end{equation}
To compute the resonance contribution we use the PDG
values~\cite{Beringer:1900zz} for the bound-state masses and leptonic
widths and the approximation $\alpha(M_{\Upsilon(NS)}) \sim 1.036
\,\alpha$~\cite{Jegerlehner:2011mw} to relate the running QED coupling to the fine structure
constant.

The continuum contribution is evaluated by integrating over experimental
data~\cite{Aubert:2008ab} corrected for initial-state
radiation~\cite{Chetyrkin:2009fv} up to $\sqrt{s} = 11.2062\,$GeV and
assuming a flat value of $R_b = 0.3\pm 0.2$ for higher energies.

\subsection{Theory moments}
\label{sec:M_th}

In PNRQCD, the normalised cross section up to NNNLO is given by the master
formula~\cite{Beneke:2013jia}
\begin{equation}
  \label{eq:R_b}
  R_b = 12 \pi e_b^2 \im\bigg[\frac{2 N_c}{s}\bigg(c_v\bigg[c_v -\frac{E}{m_b}
  \frac{d_v}{3}\bigg]G(E)+ \dots\bigg)\bigg]\,,
\end{equation}
where $e_b$ and $m_b$ are the electric charge and pole mass of the
bottom quark. The kinetic energy $E$ is related to the center-of-mass
energy via $E = \sqrt{s} - 2\*m_b$. The Wilson coefficients $c_v$ and
$d_v$ are obtained by matching the spatial components of the
relativistic vector current to non-relativistic currents:
\begin{equation}
  \label{eq:j_matching}
  j^i = c_v \psi^\dagger \sigma^i\chi + \frac{d_v}{6m_b^2}\psi^\dagger
  \sigma^i{\bf D}^2\chi+\dots\,.
\end{equation}
$G(E)$ is the correlator of the non-relativistic current $\psi^\dagger
\sigma^i\chi$. Its poles at $E=E_N$ can be interpreted as S-wave bound
states; the behaviour near a pole is given by
\begin{equation}
  G(E) \xrightarrow{E \to E_N} \frac{|\psi_N(0)|^2}{E_N - E - i\epsilon}\,,
\end{equation}
where $\psi_N(0)$ is the wave function at the origin. The theory moments
can then be written as
\begin{equation}
  \label{eq:Mn_th}
    {\cal M}_n^\text{th} = \frac{12\pi^2 N_c e_b^2}{m_b^2}
    \sum_{N=1}^\infty\frac{Z_N}{(2m_b + E_N )^{2n+1}}
  + \int_{4 m_b^2}^\infty ds\,\frac{R_b(s)}{s^{n+1}}
\end{equation}
with the residues
\begin{equation}
  \label{eq:Zn_def}
Z_N = \frac{4\*m_b^2}{s_N}\*c_v\bigg[c_v -\frac{E_N}{m_b}
  \frac{d_v}{3}\bigg] |\psi_N(0)|^2+ \dots\,,\qquad s_N = (2\*m_b + E_N)^2\,.
\end{equation}
According to our power counting (eq.~(\ref{eq:PNRQCD_power_counting}))
the prefactors $1/s, 1/s_N$ in eqs.~(\ref{eq:R_b}), (\ref{eq:Zn_def})
could be expanded in $E,E_N \ll m_b$. We find, however, that keeping
them in unexpanded form leads to a somewhat better consistency of the
mass values extracted from different moments with $n\approx 10$. The
difference between the two approaches is within our estimate for
the perturbative error.

For the bound-state energies and residues we directly adopt the known NNNLO
results~\cite{Kniehl:1999ud,Kniehl:2002br,Penin:2005eu,Beneke:2005hg,Beneke:2007pj,Marquard:2014pea}. The
continuum Green function at NNNLO has mostly been considered in the context
of top-pair
production~\cite{Beneke:2005hg,Beneke:2014,Beneke:2008cr,Beneke:2015kwa}
and the expressions have to be modified to allow a numerical evaluation
in the limit of a vanishing width~\cite{Beneke:2014pta}. In addition to
the higher-order QCD corrections we also take into account the leading
QED contributions and effects due to a non-zero charm-quark mass up to
NNLO~\cite{Melles:1998dj,Melles:2000dq,Hoang:2000fm,Beneke:2014pta}.

\section{Quark mass determination}
\label{sec:mb_det}

To extract the bottom quark mass, we first determine a numerical value
for the pole mass via NNNLO conversion~\cite{Beneke:2005hg} from the
input mass in the potential-subtracted (PS) scheme~\cite{Beneke:1998rk} and
then compute the theory moments from (eq.~(\ref{eq:Mn_th})). This
corresponds to the PS-shift prescription introduced in~\cite{Beneke:2014}.

\begin{figure}
  \centering
  \begin{tabular}{cc}
\includegraphics[width=0.45\linewidth]{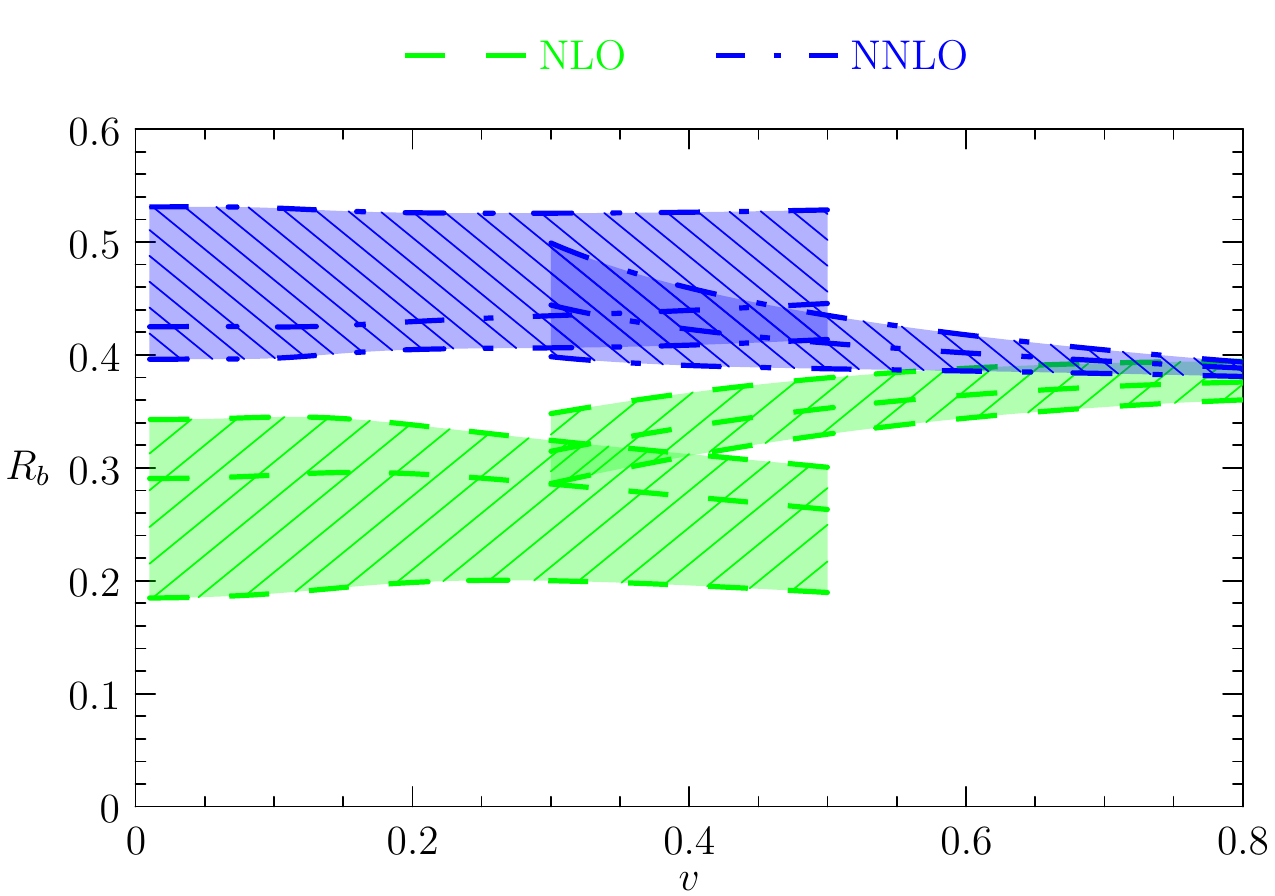}&\includegraphics[width=0.45\linewidth]{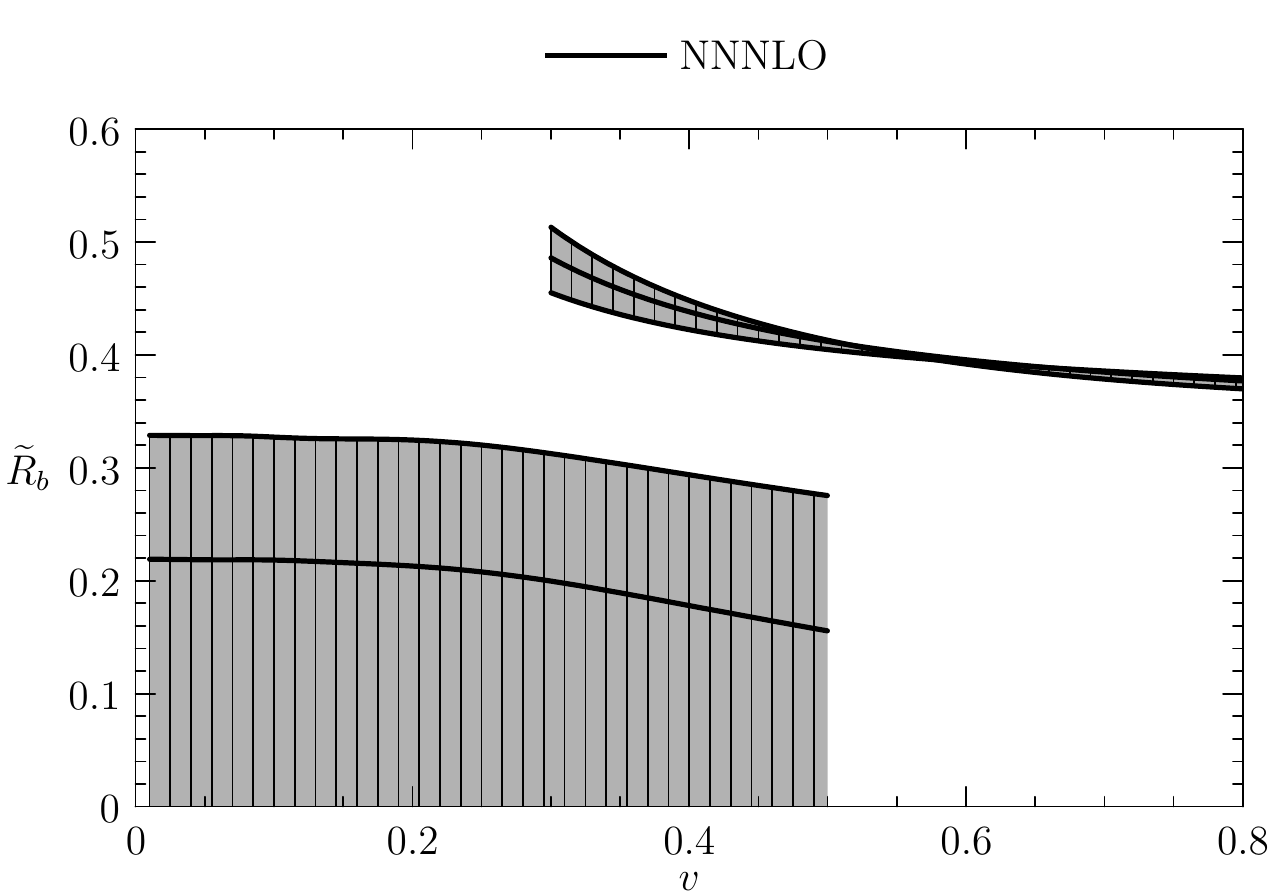}
\end{tabular}
  \caption{Behaviour of the continuum cross section as a function of $v
= \sqrt{E/m_b}$ for $m_b = 5\,$GeV. The curves on the left show the
PNRQCD prediction, whereas the curves on the right are a Pad\'e
approximation to fixed-order perturbation
theory~\cite{Hoang:2008qy,Kiyo:2009gb}. The shaded areas arise from
varying the renormalisation scale between $3\,$GeV and $10\,$GeV.}
  \label{fig:R_b_cont}
\end{figure}

\begin{figure}
  \centering
  \begin{tabular}{cc}
\includegraphics{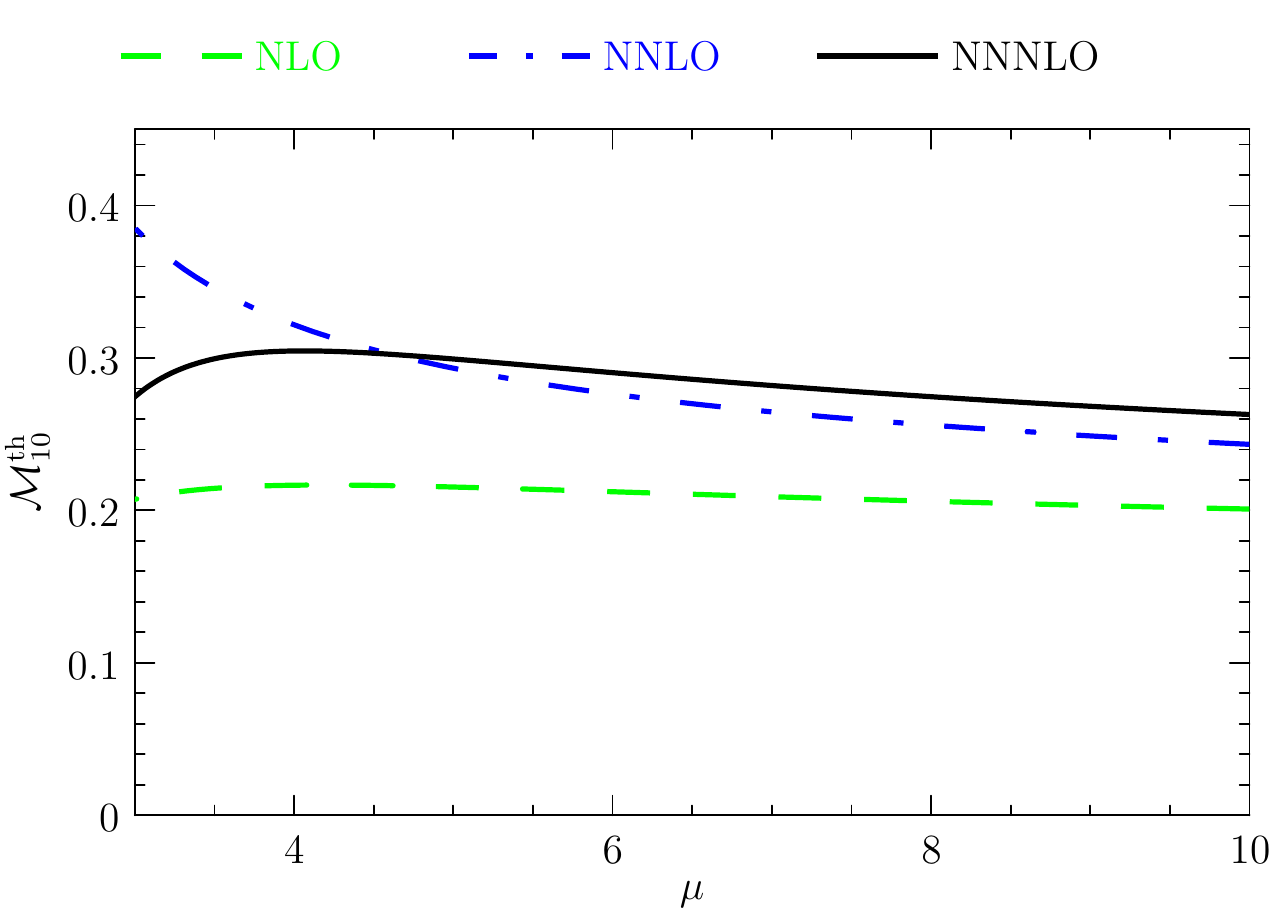}
\end{tabular}
  \caption{The 10th moment in units of $(10\,\text{GeV})^{-20}$ as a
function of the renormalisation scale with
$m_b^{\text{PS}}(2\,\text{GeV}) = 4.5\,$GeV.}
  \label{fig:M10}
\end{figure}

For small renormalisation scales $\mu \lesssim 3\,$GeV we observe no
convergence of the perturbative series, which motivates the choice of
$\mu = m_b^{\text{PS}}$ as our central scale, and perform a variation
within $3\,\text{GeV} \leq \mu \leq 10\,\text{GeV}$ to estimate the
perturbative uncertainty. Even at these comparatively high scales the
continuum cross section (cf. figure~\ref{fig:R_b_cont}) shows poor behaviour. In
particular, there is no clear convergence when going to higher orders,
and the NNNLO prediction is incompatible with the fixed-order results
for intermediate velocities. Nevertheless, the moments themselves
(figure~\ref{fig:M10}) receive only a small contribution from the continuum and
appear to be well-behaved as long as $n$ is sufficiently
large. Furthermore, the continuum contribution reduces the residual
scale dependence of the moments~\cite{Beneke:2014pta}.  This suggests
that we can indeed extract precise values for the bottom quark mass from
moments ${\cal M}_n$ with $n\approx 10$.

Our main uncertainties for the mass in the PS scheme are due to the
perturbative error, estimated as described above, the spread of the mass
values extracted from different moments $ 8 \leq n \leq 12$, and the
variation of $\alpha_s(M_Z) = 0.1184\pm 0.010$. Since we only take into
account the first six resonances in the determination of the theory
moments (eq.~(\ref{eq:Mn_th})), we assign an additional error equal to
the difference to the mass value extracted from only four resonances. To
estimate the error in the conversion from the PS to the pole scheme we
extract $m_b^{\text{PS}}$ at some intermediate scale $1\,\text{GeV} \leq
\mu_f \leq 3\,\text{GeV}$ and evolve the result to $\mu_f = 2\,$GeV. For
the experimental error we add in quadrature the uncertainties in the
$\Upsilon$ masses and leptonic widths, the uncertainty of the available
continuum data, and our estimate $0.1 \leq R_b \leq 0.5$ for high
energies.

In contrast to~\cite{Hoang:2000fm}, we find only small corrections due
to a non-zero charm-quark mass. We estimate the error from unknown
corrections beyond NNLO to be equal to the total charm-mass effect in
the bottom-quark mass determination at NNLO. We find very small QED and
non-perturbative corrections of less than $1\,$MeV and neglect the
corresponding errors.

\begin{figure}
  \centering
\includegraphics{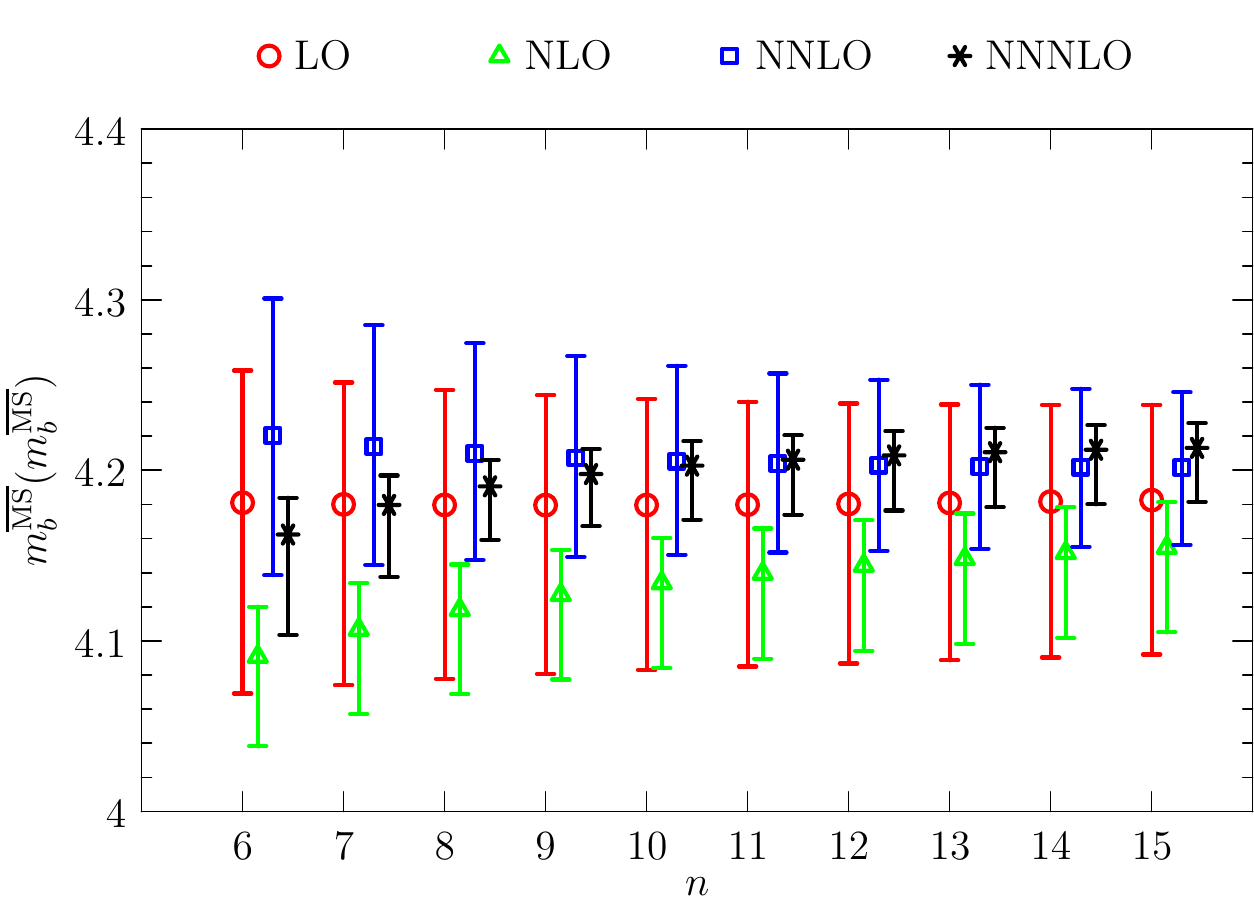}
  \caption{Values for the \MS{} quark mass $m_b^{\MS}(m_b^{\MS})$ in GeV
extracted from different moments ${\cal M}_n$ using the PS-shift prescription.}
  \label{fig:results}
\end{figure}

As our final result we adopt the bottom-quark mass at a scale of $\mu_f =
2\,$GeV determined from the 10th moment. Including the aforementioned
uncertainties we obtain
\begin{align}
  \label{eq:PS_mass_res}
  m_b^\text{PS}(2\,\text{GeV}) ={}& \big[4.532^{+0.002}_{-0.035}(\mu)
\pm 0.010 (\alpha_s) ^{+ 0.003}_{-0} (\text{res}) \pm 0.001
(\text{conv}) \notag\\
 &\pm 0.002 (\text{charm}) ^{+ 0.007}_{-0.013} (n) \pm 0.003
(\text{exp})\big]\,\text{GeV}\notag\\
={}&4.532^{+0.013}_{-0.039}\,\text{GeV}\,.
\end{align}
From our results for the masses in the PS scheme, we obtain \MS{} masses
$m_b^{\MS}(\bar{\mu})$ using four-loop scheme
conversion~\cite{Marquard:2015qpa} expressed in terms of the strong
coupling constant $\alpha_s^{(4)}(\bar{\mu})$ with four active quark
flavours. The \MS{} mass scale $\bar{\mu}$ is varied independently
between $3$ and $10\,$GeV and the resulting mass is then evolved to
$\bar{\mu} = m_b^{\MS}$. Our results for the \MS{} mass including
uncertainties are shown in figure~\ref{fig:results}. For the 10th moment
at NNNLO we obtain
\begin{align}
\label{eq:MSbar_mass_res}
  m_b^{\MS}(m_b^{\MS}) ={}& \big[4.203 ^{+ 0.002}_{-0.031} (\mu) \pm
  0.002 (\alpha_s) ^{+ 0.003}_{-0} (\text{res}) ^{+ 0.013}_{-0.004}
  (\text{conv})\notag\\
  &\pm0.002(\text{charm}) ^{+ 0.006}_{-0.012} (n) \pm 0.003
  (\text{exp})\big]\,\text{GeV}\notag\\
  ={}&4.203^{+0.016}_{-0.034}\,\text{GeV}\,.
\end{align}

As an alternative to the aforementioned PS-shift prescription, we can
use the \MS{} mass $m_b(\bar{\mu})$ instead of the PS mass to determine a
numerical value for the pole mass used in the computation of the
moments. This defines the \MS{}-shift prescription. In analogy to the
PS-shift treatment, we choose a central renormalisation scale $\mu =
m_b(\bar{\mu})$ and use $\alpha_s^{(4)}(\mu)$ both for the scheme
conversion and the calculation of the moments. We arrive at
\begin{align}
\label{eq:MSbar_mass_shift}
  m_b^{\MS}(m_b^{\MS}) ={}& \big[4.204 ^{+ 0.000}_{-0.019} (\mu) \pm
  0.002 (\alpha_s) ^{+ 0.003}_{-0} (\text{res}) ^{+ 0.002}_{-0.005}
  (\text{conv})\notag\\
  &\pm0.002(\text{charm}) ^{+ 0.007}_{-0.013} (n) \pm 0.003
  (\text{exp})\big]\,\text{GeV}\notag\\
  ={}&4.204^{+0.008}_{-0.024}\,\text{GeV}\,.
\end{align}
While the central value is in excellent agreement with
eq.~(\ref{eq:MSbar_mass_res}), the scale uncertainty in
eq.~(\ref{eq:MSbar_mass_shift}) is considerably smaller. In fact, it
does not cover the central value $m_b^{\MS}(m_b^{\MS}) = 4.177\,$GeV we
obtain when expanding the prefactors $1/s, 1/s_N$ in
eqs.~(\ref{eq:R_b}), (\ref{eq:Zn_def}) as discussed in
section~\ref{sec:M_th}. Since the preference for the unexpanded
prefactors is not based on systematic considerations, as discussed
in~\cite{Beneke:2014pta}, we quote the estimate
eq.~(\ref{eq:MSbar_mass_res}) as our final result.

Compared to~\cite{Beneke:2014pta}, our results for the \MS{} mass are
shifted upwards by $10\,$MeV and the uncertainty from the scheme
conversion is reduced significantly. These changes are due to the
recently calculated value for the four-loop coefficient in the scheme
conversion~\cite{Marquard:2015qpa}, which is smaller by about 8\%
compared to the estimate~\cite{Ayala:2014yxa} used
in~\cite{Beneke:2014pta}.

\section*{Acknowledgements}
\label{ack}

This work has been supported by the DFG
Sonder\-forschungs\-bereich/Transregio~9 ``Com\-pu\-ter\-gest\"utzte
Theoretische Teil\-chen\-physik'', the Gottfried Wilhelm Leibniz
programme of the Deutsche Forschungsgemeinschaft (DFG), and the Munich
Institute for Astro- and Particle Physics (MIAPP) of the DFG cluster of
excellence "Origin and Structure of the Universe". A.~M. is supported by
a European Union COFUND/Durham Junior Research Fellowship under EU
grant agreement number 267209.


\bibliographystyle{JHEP}
 \bibliography{biblio}

\end{document}